\documentclass[article]{aa}

\usepackage{txfonts}
\usepackage{graphicx}
\usepackage{natbib}
\bibpunct{(}{)}{;}{a}{}{,}

\begin{document}

\title{First stars and the extragalactic background light: How recent $\gamma$-ray observations constrain the early universe.}

\author{Martin Raue\inst{1}
  \and Tanja Kneiske\inst{2}
   \and Daniel Mazin\inst{3}}
  
\institute{Max-Planck-Institut f\"ur Kernphysik, Saupfercheckweg 1, 69117 Heidelberg, Germany
\and Institut f\"ur Experimentalphysik, Universit\"at Hamburg, Luruper Chaussee 149, 22765 Hamburg, Germany
\and Institut de Fisica d'Altes Energies (IFAE), Edifici Cn. Universitat Autonoma de Barcelona, 08193 Bellaterra (Barcelona), Spain}
  
\offprints{M. Raue, \email{martin.raue@mpi-hd.mpg.de}}

\date{Accepted for publication in A\&A}

\abstract
{The formation of the first stars (Population III; PopIII) marks the end of the dark ages of the universe, a subject of lively scientific debate. Not (yet) accessible to direct observations, this early stage of the universe is mostly studied via theoretical calculations and numerical simulations. An indirect window is provided by integrated present day observables such as the metal abundance or the diffuse extragalactic photon fields.}
{We aim to derive constraints on the properties of the
PopIII and low metallicity Population II (LM PopII)
stars utilizing limits on the density of the extragalactic background light (EBL), recently derived from very-high-energy (E$>$100\,GeV; VHE) observations.} 
{A model calculation for the evolving EBL density produced by
PopIII/LM PopII
stars is presented. The model utilizes stellar population spectra (SPS) for zero and low metallicity stars and accounts for the changing emission of an aging stellar population. Emission from the dense HII regions surrounding the stars (nebula) is included. The resulting EBL density for different scenarios (metallicity, star formation rate, initial mass function) is compared to the limit on the EBL density. The potential for detecting a cut-off in HE/VHE spectra is discussed.} 
{Assuming a maximum contribution from
PopIII/LM PopII stars to the EBL density
of 5\,nW\,m$^{-2}$\,s$^{-1}$ at 2\,$\mu$m a limit on the star formation rate (SFR) of the first stars of 0.3 to 3\,M$_\odot$\,Mpc$^{-3}$\,yr$^{-1}$ in the redshift range $7 - 14$ is derived. The limit depends on the assumed shape of the SFR and metallicity.}
{
The EBL can be used as a probe to investigate the properties of
PopIII/LM PopII stars.
Limits on the EBL density derived from VHE observations can provide constraints on the parameters of the these stars, in particular the star formation rate.}


\titlerunning{First stars and the extragalactic background light}

\maketitle

\section{Introduction}\label{Sec:Introduction}

The end of the dark ages of the universe - the epoch of reionization - is a field of great interest (e.g. \citealt{barkana:2001a,ciardi:2005a}). This epoch is associated with the formation of the first stars (Population III; PopIII)%
\footnote{See \citet{oshea:2008a} for a discussion on naming conventions.}  (e.g. \citealt{bromm:2004a,glover:2005a}),
which are believed to start the reionization of the universe at redshift of about $z= 10 - 30$.
PopIII stars form in a pristine environment, in clouds of hydrogen and helium with little or no heavy elements (primordial composition). Due to the absence of heavy elements, the cooling of such collapsing gas clouds is likely dominated by H$_2$ cooling through molecular emission lines. Numerical simulations of collapsing clouds with primordial composition predict very massive stars (100-1000\,M$_\odot$) with high effective temperatures  ($\sim 10^{5}$\,K) and short lifetimes ($\sim10^6$\,yrs) (e.g. \citealt{bromm:1999a,bromm:2002a,abel:2002a}). Such hot massive stars produce copious amount of ionizing photons \citep{schaerer:2002a} and can therefore reionize the universe. The formation of lower mass stars is also possible, if e.g. the cloud cooling is driven by hydrogen-deuterium (HD) and atomic hydrogen (H) cooling \citep{uehara:2000a, johnson:2006a}. Other processes including turbulent fragmentation \citep{klessen:2005a}, magnetically-regulated fragmentation \citep{silk:2006a} and dust cooling at very high densities \citep{omukai:2005a} could also explain stars with lower masses $< 100$\,M$_\odot$%
\footnote{For a more complete discussion on formation of the first stars see e.g. the 2008 updated version of \citealt{ciardi:2005a}, astro-ph/0409018.}.

PopIII stars produce the first heavier elements, paving the way for the second generation of stars. When the star forming cloud reaches a critical metallicity ($Z_{\mathrm{CR}} \sim 10^{-6} - 10^{-4}$\,Z$_{\odot}$ e.g. \citealt{schneider:2006b,omukai:2005a}) cooling through line emission from heavier atoms (C,O) and molecules (H$_2$O, CO, O$_2$) starts to dominate. Thereby, the second generation of stars with (likely) lower masses and "Salpeter-like" initial mass function start to form (Population II; PopII) (see e.g. \citealt{bromm:2001a,schneider:2002a,schneider:2003a,bromm:2003a,schneider:2006a}). The transition from dominant PopIII to PopII star formation could already happen at early times (e.g. $z \gg 7$), since pair-instability and core-collapse supernova explosion from PopIII stars can effectively enrich their environments with metals \citep{schneider:2002a,scannapieco:2003a,bromm:2003a,tornatore:2007a}.

Direct observations of this early period of the universe are challenging: halo stars with extremely low metallicities have been detected in our galaxy \citep{christlieb:2002a}, but the observation of a true PopIII star with zero-metallicity is still pending. The upcoming satellite experiment James Webb Space Telescope (JWST)\footnote{http://ngst.gsfc.nasa.gov}, expected to be launched in 2013, with high sensitivity in the 1-10\,$\mu$m near-infrared (NIR) band is aiming to detect the redshifted ultraviolet (UV) to optical (O) emission from source at high redshifts $z > 10$.

Other constraints on the PopIII stars can be derived from integrated properties like e.g. the number of baryons bound in stars or the number of ionizing photons produced. If the contribution from other sources to these integrated properties are reasonably well known, the contribution from PopIII stars can be derived. This can then be compared with model calculations for different PopIII scenarios. \citet{tumlinson:2006a} simulated the formation of PopIII stars using galactic chemical evolution models and compared the model output with the present day metallicity distribution function (MDF) of the Galaxy. They found that, while not yet formally conclusive, the MDF could best be described by a PopIII initial mass functions which includes lower mass stars.

\cite{nagamine:2006a} used several integrated properties including the extragalactic background light (EBL; see next paragraph) to derive constraints on the cosmological star formation history of PopII stars.

In the optical to near-infrared (O-NIR) wavelength regime of the diffuse meta-galactic photon field (extragalactic background light; EBL)%
\footnote{We will use the term extragalactic background light (EBL) to denote the diffuse meta-galactic photon field in the UV to IR wavelength regime.} stars are the main contributors to the EBL density.
Luminous PopIII stars can leave a distinct signature in the EBL density  \citep{bond:1986a}. In particular, their contribution may exceed significantly the EBL density inferred from low redshift ($z < 5$) sources.
Direct measurements of the EBL are difficult due to dominant foregrounds in our planetary system (zodiacal light) and the Galaxy \citep{hauser:1998a}. Nevertheless, the discovery of such a NIR background excess (NIRBE) with high significance has been claimed by \citet{matsumoto:2005a}, while other data showed a marginal excess (see \citealt{hauser:2001a} for a review). The nature of this excess is still under debate. \citet{dwek:2005c} find that it is likely a foreground artifact from zodiacal light and not of extragalactic origin. \footnote{In addition, \citet{mattila:2006a} argued that the claimed discontinuity in the EBL at UV-O wavelengths, which has been interpreted as a signature for the first stars, is also an artifact of foreground subtraction.} A possible PopIII origin of the NIRBE has been investigated by many authors \citep{santos:2002a,salvaterra:2003a,dwek:2005c,madau:2005a,salvaterra:2006a,fernandez:2006a}. While \citet{dwek:2005c} and \citet{madau:2005a} argue that the number of stars required to produce such an excess would overproduce todays metallicity and would lead to a too high number of baryons in stars, \citet{fernandez:2006a} (FK06) find that, if accounting for the final stage of the first stars in more detail, a PopIII origin of the NIRBE seems possible.

An indirect way of deriving constraints on the EBL comes from the measurement of very high energy (VHE) $\gamma$-ray spectra from distant sources \citep{stecker:1992a}. VHE $\gamma$-rays interact with low energy photons from the EBL via pair-production \citep{nikishov:1962a,gould:1967a}. The cross-section of the pair-production is strongly peaked, so this process leaves an energy dependent attenuation signature in the measured VHE spectra. With assumption about the source physics, upper limits on the EBL density can be derived (e.g. \citealt{dwek:2005a,aharonian:2006:hess:ebl:nature,mazin:2007a}). \citet{dwek:2005b} considered the effect of a high NIRBE on the spectra of distant blazars and concluded that such a high density as reported by \citet{matsumoto:2005a} seemed unlikely. Recently, strong limits on the EBL density in the NIR have been derived (e.g. \citealt{aharonian:2006:hess:ebl:nature,mazin:2007a}), which exclude the claimed NIRBE with high significance, and are only a factor $\sim 2$ above the lower limits derived from source counts \citep{madau:2000a}.

In this paper these limits on the diffuse EBL density are used to derive constraints on the properties of the PopIII/LM PopII stars. Results from a detailed model calculation of the EBL for different PopIII/LM PopII star scenarios are compared with recent limits on the EBL density. Our model accounts for the time evolution of the emissivity of a stellar population, which, for the case of low mass stars with long lifetimes, has profound implications for the resulting EBL.

The paper is organized as follows: In Sec.~\ref{Sec:EBL_Model} the model calculations for the EBL density from  PopIII/LM PopII stars are described. In Sec.~\ref{Sec:Constrains} the resulting EBL density for different sets of  PopIII/LM PopII star parameters is calculated and compared with recent limits. Limits on cosmological star formation rate (SFR) are derived and the detectability of a cut-off in high energy spectra resulting from EBL attenuation are discussed. In Sec.~\ref{Sec:Discussion} the derived limits are compared with previous results and the consequences for the PopIII/LM PopII star properties are discussed. We summarize our results in Sec.~\ref{Sec:Summary}.

Throughout this paper flat Friedman cosmology is adopted with $\Omega_0 = 0.25$ , $\Omega_\Lambda=0.75$  and a Hubble constant of $H_0=70$~km s$^{-1}$ Mpc$^{-1}$.

\section{EBL Model}\label{Sec:EBL_Model}

Previous models for the EBL density produced by PopIII/LM PopII stars often focused on very massive stars with $M \sim 1000$\,M$_\odot$ \citep{santos:2002a,dwek:2005c,madau:2005a}. For such very massive stars, simplified assumptions can be made in the calculations, e.g., the stellar emission follows a black body spectrum, the total emission is dominated by photons reprocessed in the surrounding HII region and the lifetime is short enough to neglect the luminosity evolution. Here we want to explore more realistic scenarios for PopIII stars, which likely form with a wider range of masses. We therefore focus on a time dependent modeling of the evolving emissivity.

We follow the equations presented in \citet{kneiske:2002a} to calculate the EBL. The two basic ingredients for the calculation are (i) the cosmic star formation rate ${\rho}_{\ast}$ and (ii) the luminosity $L_{\nu}(\tau)$ of a specific stellar population of age $\tau$ as e.g. derived from stellar population syntheses models. The co-moving emissivity (luminosity density) at redshift $z$ is obtained from the convolution
 \begin{equation}
 \mathcal{E}_{\nu}(z) = \int_z^{z_{m}}
 L_{\nu}(t(z)-t'(z'))\,\rho_{\ast}(z') \left | \frac{dt'}{dz'} \right |
 dz' ~,
 \label{eq:emislambda}
 \end{equation}
 where the cosmic star formation rate ${\rho}_{\ast}(z')$ is assumed to begin at some finite epoch
 $z_m=z(t_m)$ and $t(z)/t'(z')$ is the cosmic time corresponding to a redshift $z/z'$. For given evolution of the emissivity a second integration over
 redshift yields the energy density, or, after multiplication with $c/4\pi$, the
 co-moving power spectrum of the EBL
 \begin{equation} P_\nu(z) = \nu
 I_{\nu}(z) = \nu \frac{c}{4\pi} \int_z^{z_m}  \mathcal{E}_{\nu'}(z')
\left | \frac{dt'}{dz'} \right |  dz' ~ ,
\label{eq:hinter} 
\end{equation}
 with $\nu'=\nu(1+z')/(1+z)$. Further details on the calculation can be found in \citet{kneiske:2002a}.

Photons from the star ionize the dense gas cloud surrounding the formation site. This HII region will re-emit photons in emission lines, free-free and free-bound, and two-photon emission (nebula emission). For the hot and massive stars the nebula emission can become dominant (e.g. \citealt{schaerer:2002a}). Therefore, two contributions to the specific luminosity are considered $L_{\nu}(t)$:
\begin{equation}
L_{\nu}(t) = L_{\nu}^{\mathrm{stars}}(t) + L_{\nu}^{\mathrm{nebula}}(t) \, ,
\end{equation}
with $L_{\nu}^{\mathrm{stars}}(t)$ the stellar emission and $L_{\nu}^{\mathrm{nebula}}(t)$ the nebula emission.
Since we are interested in an averaged and integrated property, the EBL, we keep the model as simple as possible only considering dominant contribution to the overall emission.

\subsection{Stellar emission}

Massive PopIII stars can effectively enrich their environment with metals via supernova explosions (see Sec.~\ref{Sec:Introduction}). It is therefore possible, that already at early times the total stellar emission is dominated by the emission from PopII stars with low metallicity. To probe the effect of these two scenarios, we will use two stellar populations with different metallicities for the EBL calculation: (1) stars with zero metallicity corresponding to dominant PopIII star emission and (2) stars with low metallicity corresponding to dominant PopII star emission. In reality, of course, the transition between PopIII and PopII stars is likely an extended and patchy process, with parallel formation of stars in zero and metal enriched environments   (e.g. \citealt{tornatore:2007a}). Here we focus on just the two extreme scenarios to identify possible differences and observational signatures.

\subsubsection{Primordial/Zero metallicity (ZM)}

\begin{figure}[t,b]
\centering
\includegraphics[width=0.5\textwidth]{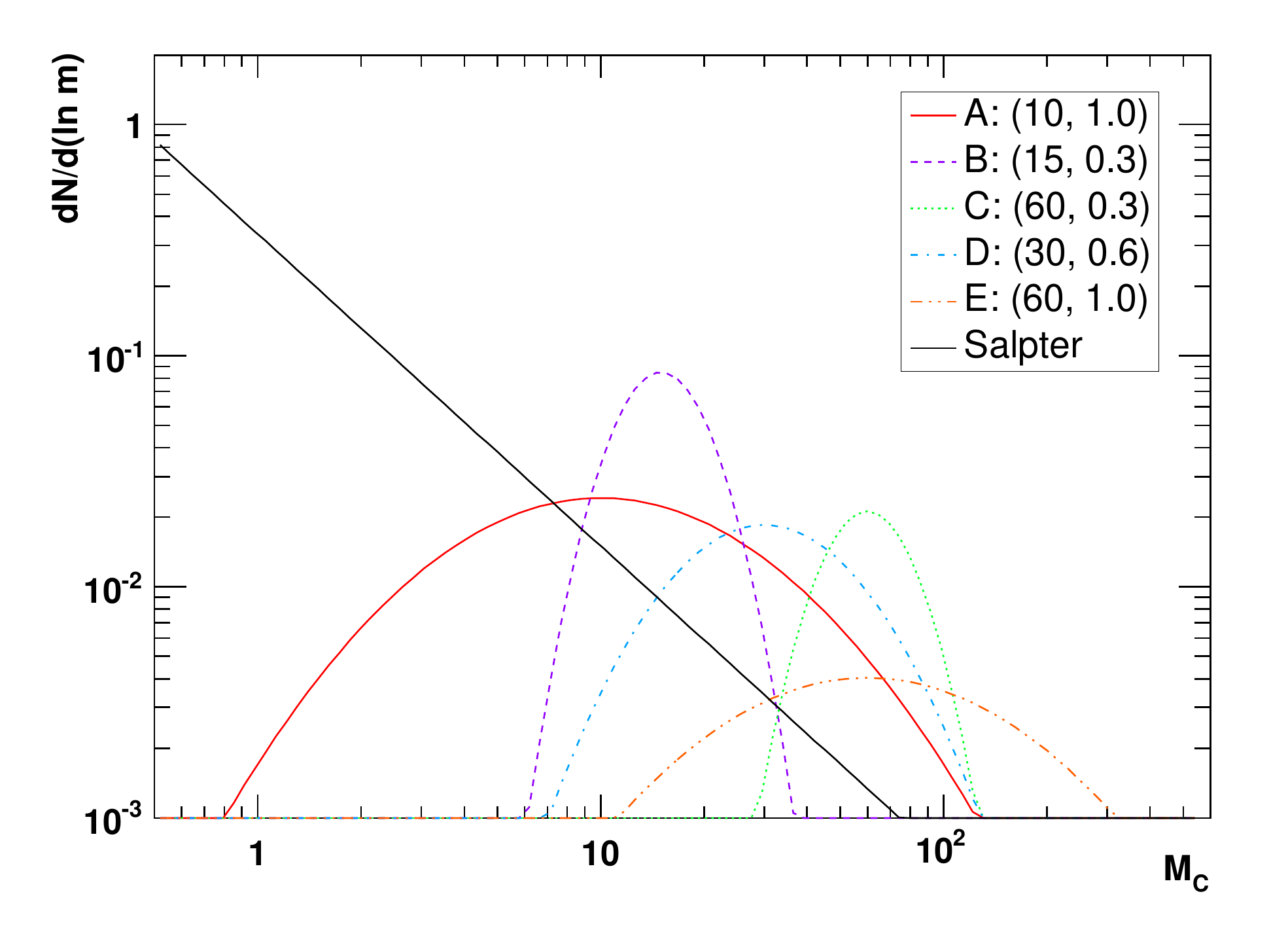}
\caption{Normalized stellar initial mass functions (IMFs) for the IMF test cases A to E from \citet{tumlinson:2006a}. All Tumlinson IMFs follow a lognormal distribution with peaks in the range between 10 and 80\,$M_{\odot}$. The values in brackets in the legend denote the peak position and the width of the lognormal distribution. For comparison a power law IMF with Salpeter slope $\alpha = 1.35$ is also shown.}
\label{Fig:TumIMFs}
\end{figure}
 
\begin{figure}[t,b]
\centering
\includegraphics[width=0.5\textwidth]{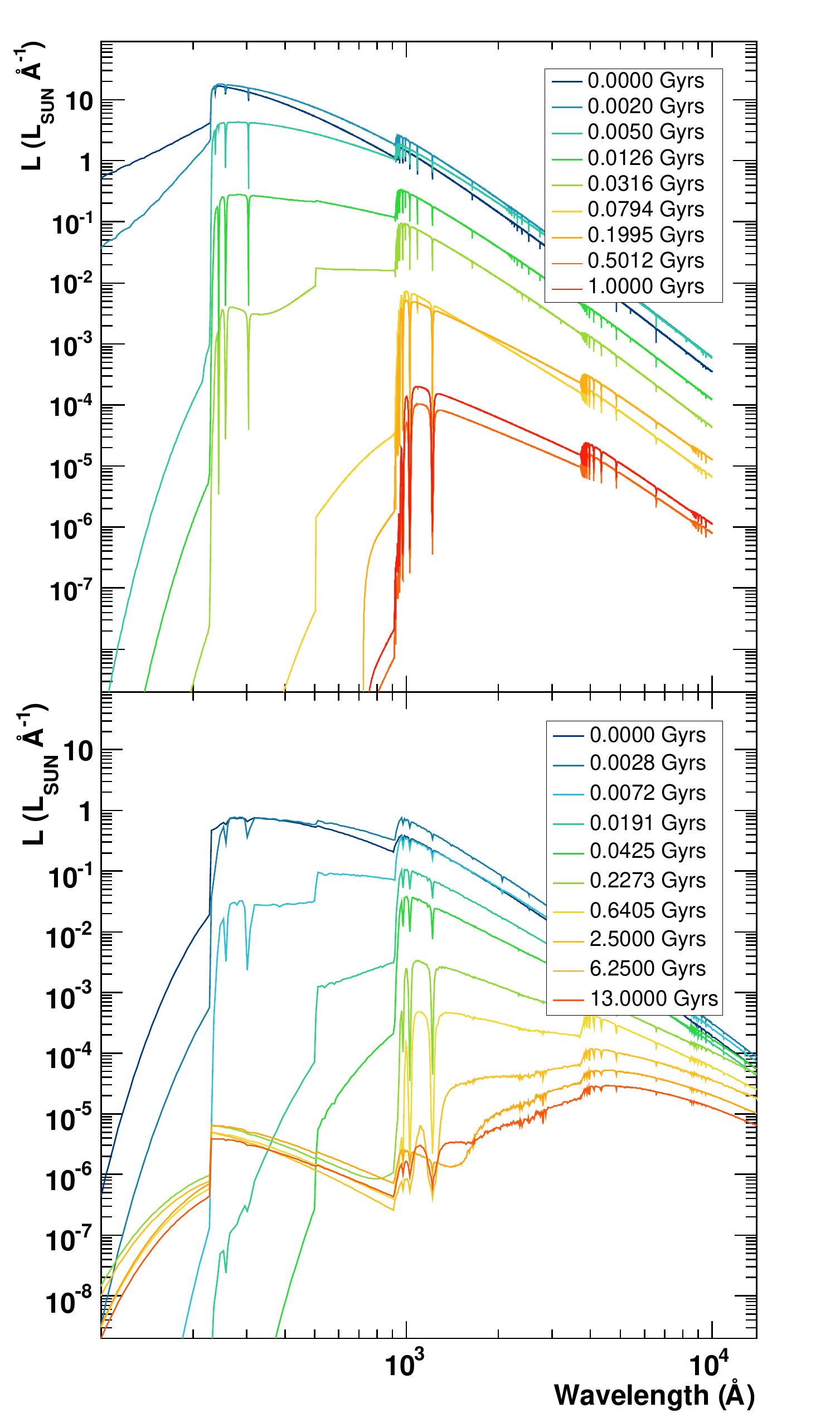}
\caption{Spectra of stellar populations as function of age $\tau$. \textit{Top:} Primordial/Zero metallicity TumA from \citet{tumlinson:2006a}; $\tau$ ranges from 0 to 1\,Gyr (top to bottom). \textit{Bottom:} Low metallicity $Z=10^{-4}$ from \citet{bruzual:2003a}; $\tau$ ranges from 0 to 13\,Gyr (top to bottom).}
\label{Fig:SPS_TumA_BC}
\end{figure}
 
Emissivities for primordial/zero metallicity (ZM) stars are based on the stellar models from \citet{tumlinson:2003a} and \citet{tumlinson:2006a}. \citet{tumlinson:2003a} solved the coupled stellar equations for ZM stars with a variant of the relaxation method taking into account proton-proton, CNO and He burning for energy production. The stellar atmospheres have been simulated with the TLUSTY code. Stellar population spectra (SPS) for a synthetic stellar cluster have been constructed from evolutionary tracks. \citet{tumlinson:2006a} connected present day observational data, like the galactic metal distribution function, with properties of the PopIII stars via galaxy chemical evolution models. The author tested a wide range of different initial mass functions (IMF; all lognormal, see e.g.  \citealt{larson:1973a}) and constructed five IMF test cases labeled A to E (TumA-E), which are shown in Fig.~\ref{Fig:TumIMFs}. All IMFs test-cases are top heavy with peaks in the range between 10 and 80\,$M_{\odot}$ and different widths. We will adopt the TumA case as our fiducial model and discuss the differences in the EBL resulting from the different IMFs in Sec.~\ref{SubSec:ConstraintsIMF}. The SPSs for the case TumA and different ages are shown in Fig.~\ref{Fig:SPS_TumA_BC} top panel. Nuclear burning in ZM stars occurs at higher temperatures than in PopII stars \citep{tumlinson:2003a}, which results in a large number of H and He ionizing photons in the UV. High mass stars have a short lifetime, so the overall emission of a high mass stellar population is significantly reduced at stellar population ages $\geqslant 0.1$\,Gyrs.

\subsubsection{Low metallicity (LM)}

For stellar populations with low metallicity (LM) the well established Isochrone
Synthesis Spectral Evolutionary Code from \citet{bruzual:2003a} is utilized.
This code has also been used for the PopII star component in the EBL model of  \citet{kneiske:2002a}. Here we choose a metallicity of $Z=10^{-4}$ and a Salpeter IMF. In contrast to the lognormal IMFs of the ZM stars described in the previous section, this will lead to a large number of stars with masses $<10$~M$_\odot$ (Fig.~\ref{Fig:TumIMFs}). The SPSs for different ages are shown in Fig.~\ref{Fig:SPS_TumA_BC} bottom panel. Due to the longer lifetime of the low mass stars, the overall emission time of the LM population is significantly longer than in the case of ZM stars with high masses.

The LM second generation stars already form in (low) dusty environments. The dust formation at such high redshifts is not known and strongly depends on the environment. Dust re-emission would occur in the MIR to FIR and redshifts into the microwave regime, which is completely dominated by the cosmic microwave background (CMB). For the wavelength range considered here (O-NIR) dust absorption is negligible and we will not consider it further.

\subsection{Nebula and line emission}

\begin{figure}[t,b]
\centering
\includegraphics[width=0.45\textwidth]{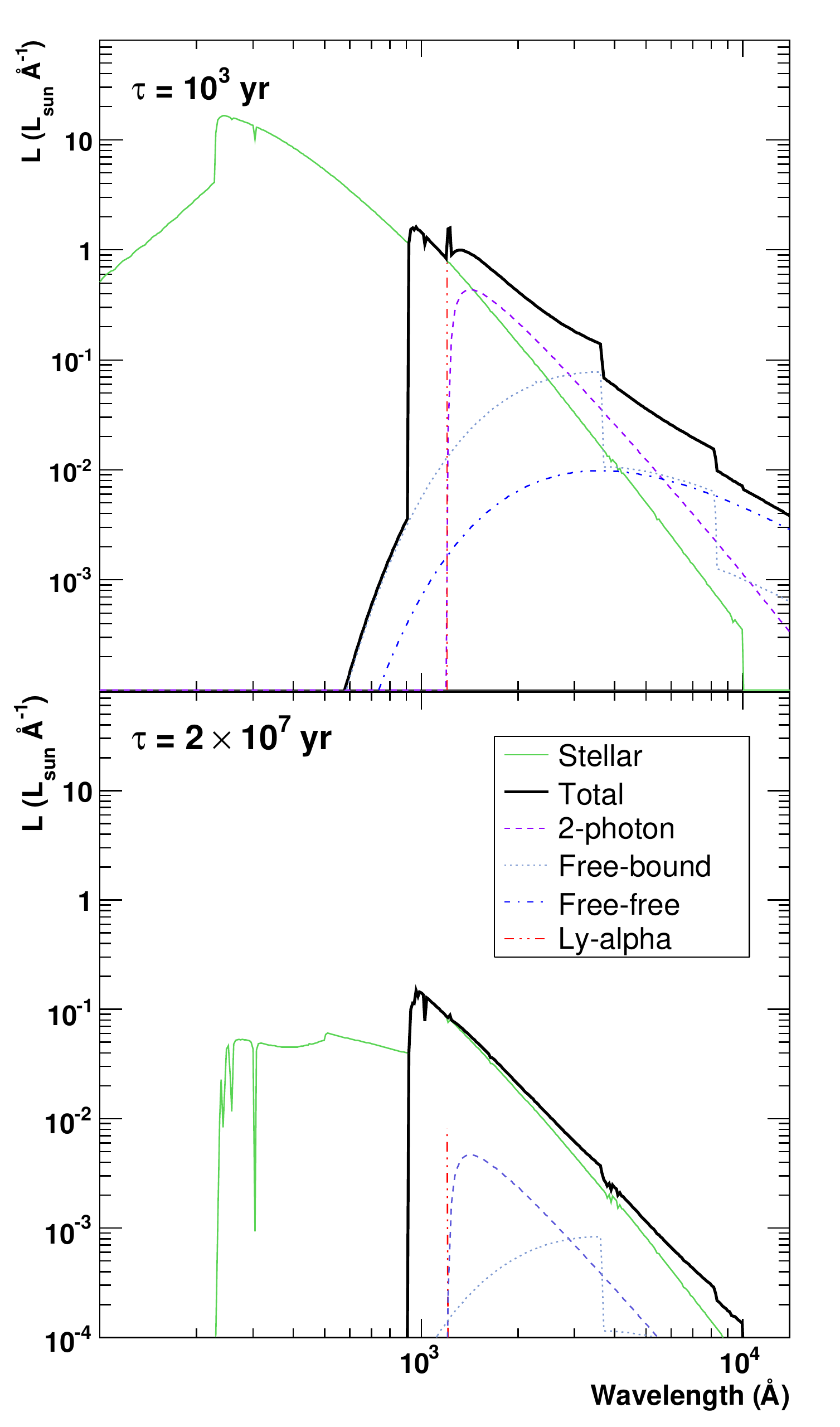}
\caption{Stellar and nebula emission for stellar population of age $10^{3}$ (upper panel) and $2 \times 10^{7}$ years (lower panel) (stellar emission: ZM, TumA). The emission components shown are stellar emission (green solid), Lyman-$\alpha$ line (red dashed double-dotted), 2-photon (purple dashed), free-bound (light-blue dotted), and free-free continuum emission (dark-blue dashed-dotted). The thick solid line is the total emission.}
\label{Fig:SPS_TumA_Nebula}
\end{figure}
 
The PopIII stars are embedded in dense gas clouds, which absorb and re-process photons from the star, creating an H\,II region (nebula). They burn at higher temperatures than their successors \citep{tumlinson:2003a}, producing copious amounts of H and He ionizing photons. The nebula absorption and emission can therefore severely alter the SPS spectra \citep{santos:2002a, schaerer:2002a,fernandez:2006a}.

In our calculations, we assume total absorption of hydrogen-ionizing photons in the nebula.
As discussed in \citet{fernandez:2006a}, the continuum luminosity does not depend on the number density of electrons or protons in the nebula, since, in case of a Str\"omgren sphere, a higher number densities would result in a higher re-combination rate and consequently a smaller emission region. This approximation is correct as long as the bulk of the emission comes from the nebula around the star in the host halo.
The effect of escaping UV photons into the inter-galactic medium (IGM) on the SPS spectrum has been calculated by \citet{santos:2002a}. The absorption of escaping ionizing photons would then take place in the IGM leading to a similar spectrum in the O-NIR wavelength range as in the case of absorption in the nebula.\footnote{At larger wavelengths, the spectra are different due to the different strength in the free-free continuum emission.}

To keep the calculation simple, we will only consider the continuum and line emission of hydrogen, following the calculations of  \citet{fernandez:2006a} (FK06 in the following). The contributions from free-free and free-bound continuum emission, 2-photon emission, and the Lyman-$\alpha$ line emission are taken into account. Below, the equations governing the nebula emission are briefly summarized. For details we refer the reader to FK06.

The luminosity of the free-free and free-bound continuum emission is given by
\begin{equation}
L^{cont}_{\nu} = \frac{\epsilon_\nu Q_H}{n_e n_p \alpha}
\end{equation}
where $Q_H=\int_{\nu_{ion}}^\infty L^{star}_{\nu} \nu^{-1}\mathrm{d} \nu$  is the production rate of hydrogen ionizing photons, $n_e$ and $n_p$
number density of electrons and protons, respectively, 
$\alpha_B\approx 2.06\cdot10^{-11}T^{-1/2}_g$ cm$^3$ s$^{-1}$ is the case-B recombination
coefficient for hydrogen at a temperature of $T_g=20.000$~K (see \citealt{spitzer:1978a}).

The total emissivity for free-free and free-bound emission taken from \citet{dopita:2003a} is
\begin{equation}
\epsilon_\nu = 4\pi n_e n_p \gamma_c \frac{\exp(-h\nu/kT_g)}{T^{1/2}_g}
\end{equation}
The coefficient $\gamma_c$ is a constant $F_k = 5.44\cdot 10^{-39}$ times
a term for free-free and free-bound emission
\begin{equation}
\gamma_c = F_k(\langle g_{ff} \rangle+\sum_{n-n'}^\infty \frac{x_n \exp{x_n}}{n} \langle g_{fb} \rangle)
\end{equation}                              
with $x_n=Ry/(kT_g n^2)$ and $ \langle g_{ff} \rangle \approx1.1$ and $\langle g_{fb} \rangle \approx1.05$ are the gaunt factors for
free-free and free-bound emission, respectively (FK06).
Note that for free-bound emission the sum is over all bound states with energies 
below the frequency in question and the infinite sum over $n$.

The luminosity for the line contribution is given by
\begin{equation}
L^{line}_\nu = \sum_i h\nu_i \phi_i(\nu-\nu_i)f_iQ_H
\label{eq:lumi_line}
\end{equation}
where $\phi_i(\nu-\nu_i)$ is the line profile which is assumed
to be a delta function. The fraction of ionizing photons which are converted to a line i is 
$f_i$. To keep the model simple only the Lyman-$\alpha$ line emission will be included, 
since the effect of other series is very small \citep{santos:2002a,dwek:2005c}.
A fraction of $f_{Ly\alpha}=0.66$ is used (see \citealt{spitzer:1978a} for 20.000K),
because 2/3 of the time a n=2 $\rightarrow$ 1 transition creates a Lyman-$\alpha$ photon
and 1/3 of the time a continuum photon via 2-photon decay is emitted.

The luminosity of the two-photon emission is calculated by
\begin{equation}
  L_\nu^{2\gamma}
  = \frac{2h\nu}{\nu_{Ly_\alpha}}(1-f_{Ly_\alpha})P(\nu/\nu_{ly\alpha})Q_H
\end{equation}
where P(y)dy is the normalized probability per 2-photon decay of getting one photon 
in the range $dy=d\nu/\nu_{Ly_\alpha}$. An analytical expression for  $P(y)$ has been derived from a polynomial fit to the data given in Table 4 of \citet{brown:1970a}, which also fits well the data for $y > 0.5$ (e.g. \citealt{gaskel:1980a}):
\begin{eqnarray}
  P(y) & = & 1.307 - 2.627(y-0.5)^2 + 2.563(y-0.5)^4 \\
  & & - 51.69(y-0.5)^6. \nonumber
\end{eqnarray}

The different components of the nebula emission together with the stellar and total emission for a stellar population of age  $10^{3}$  and $2 \times 10^{7}$ years are shown in Fig.~\ref{Fig:SPS_TumA_Nebula} (ZM stellar population model, TumA). For young stellar populations the nebula emission dominates the emission spectrum at wavelengths $\lambda \gtrsim 1200 \, \AA $. Older stellar populations produce less ionizing photons, so the nebula emission gradually decreases until it is negligible.

\subsection{Star formation rate}\label{SubSec:SFR}

\begin{figure}[t,b]
\centering
\includegraphics[width=0.45\textwidth]{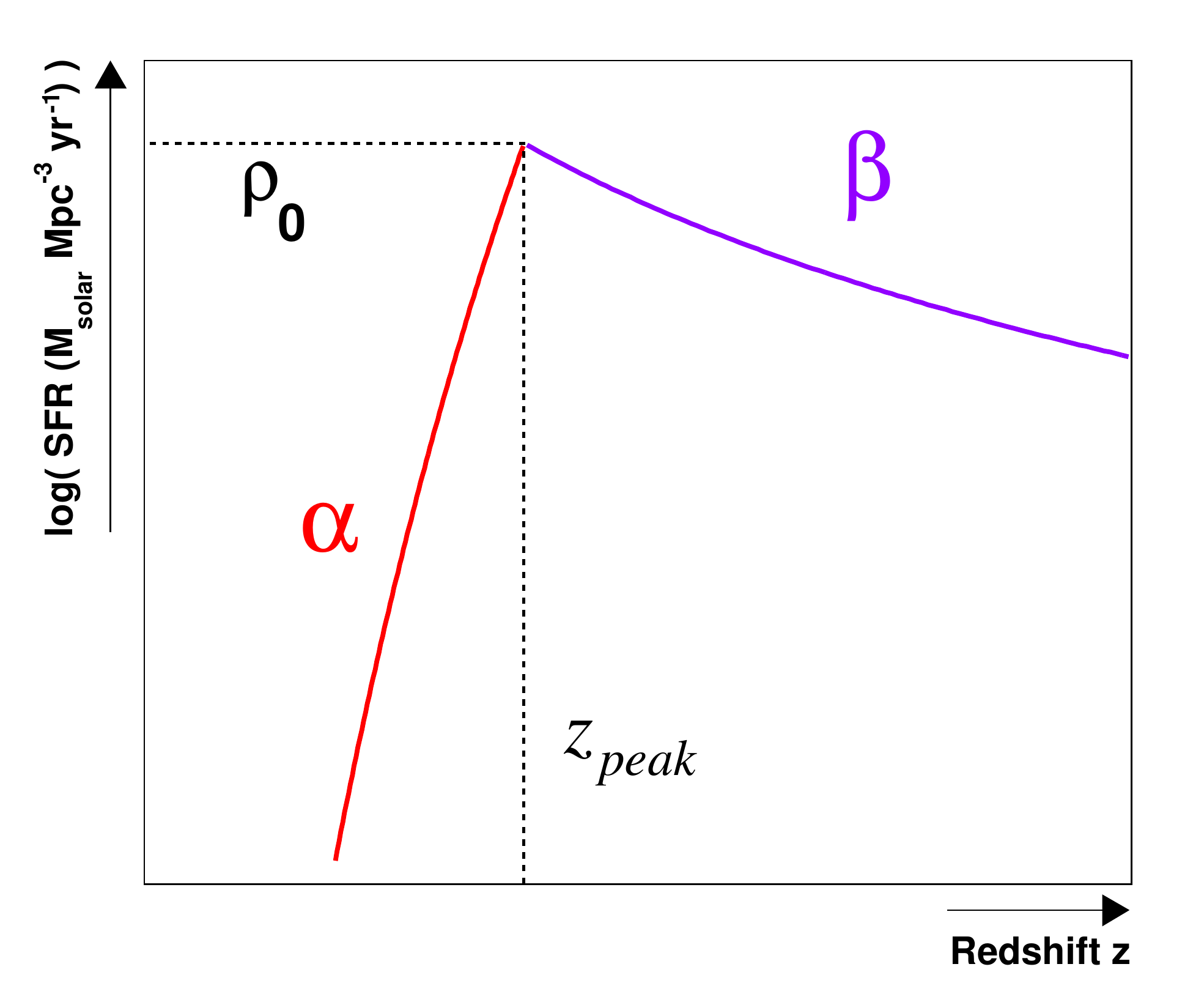}
\caption{Sketch of the parameterization for the star formation rate (SFR) as broken power-law (Eq.~\ref{Eq:SFR_BPL}). A set of SFR parameter will be denote as $\mathrm{SFR} (z_{\mathrm{peak}}, \alpha, \beta )$, with the normalization set to $\rho_{0} = 1$.}
\label{Fig:SFR_sketch}
\end{figure}
 
The star formation rate (SFR) at high redshift (primordial SFR; PSFR) is largely unknown. Direct measurements and limits on the PSFR from deep surveys of UV-bright galaxy with the dropout technique \citep{bouwens:2006a,richard:2006a,mannucci:2007a,bouwens:2007a,bouwens:2008a,richard:2008a}, Ly-$\alpha$ emitters \citep{kashikawa:2006a,ota:2008a}, and gamma-ray bursts \citep{firmani:2004a,le:2007a,guetta:2007a,yuksel:2008a} exist, but the individual measurements do not agree. While some measurements show a strong decline in the SFR at $z > 5$ (e.g. \citealt{mannucci:2007a,bouwens:2008a}), others detect a constant or only slowly declining SFR (e.g. \citealt{richard:2006a, richard:2008a, yuksel:2008a}). The different methods most likely sample different populations of PSFR regions and are not directly comparable. The individual measured PSFRs therefore can be considered as a lower limit to the total PSFR \citep{yuksel:2008a}. Furthermore, PopIII stars form in small (proto-galactic) halos (e.g. \citealt{greif:2008a}), so the galaxy PSFR, as derived from UV bright galaxies, may not be representative for the PSFR.

Further constraints on the PSFR come from constraints on the reionziation history of the universe. The detection  of absorption features in the spectra from distant quasars (Gunn-Peterson trough; \citealt{gunn:1965a}) can provide direct evidence on the ionization state of the early universe. Such observations indicate that the universe was largely ionized at $z \sim 6$ \citep{fan:2006a}. Recent measurements of the fluctuation power spectrum and polarization of the CMB from the WMAP 5 year data also point toward an early and extended reionization, with the universe mostly ionized at redshift $z_{\mathrm{reion}} \sim 11 \pm 1.4$ \citep{dunkley:2008a}. While the sources of the reionization are not well known, PopIII/LM PopII stars are considered to be the natural candidates \citep{barkana:2001a}.

For our model, we adopt a simple broken power law to describe the primordial SFR:
\begin{equation}
{\rho}_{\ast}(z) = {\rho}_{0} \cdot \left ( \frac{z + 1}{z_{\mathrm{peak}} + 1} \right ) ^{q} \label{Eq:SFR_BPL}
\end{equation}
with $q = \alpha$ for $z < z_{\mathrm{peak}}$ and $q = \beta$ for $z  \geqslant z_{\mathrm{peak}}  $ (sketched in Fig.~\ref{Fig:SFR_sketch}). We only consider star formation in the redshift range from $z_{\mathrm{start}} = 35$ to $z_{\mathrm{stop}} = 5$, i.e.  ${\rho}_{\ast}(z > z_{\mathrm{start}}) = 0$ and ${\rho}_{\ast}(z < z_{\mathrm{stop}}) = 0$. In the following a set of SFR parameters will be denoted as $\mathrm{SFR} (z_{\mathrm{peak}}, \alpha, \beta )$.

\subsection{Resulting EBL}\label{SubSec:ResultingEBL}

\begin{figure*}[tb]
\centering
\includegraphics[width=\textwidth]{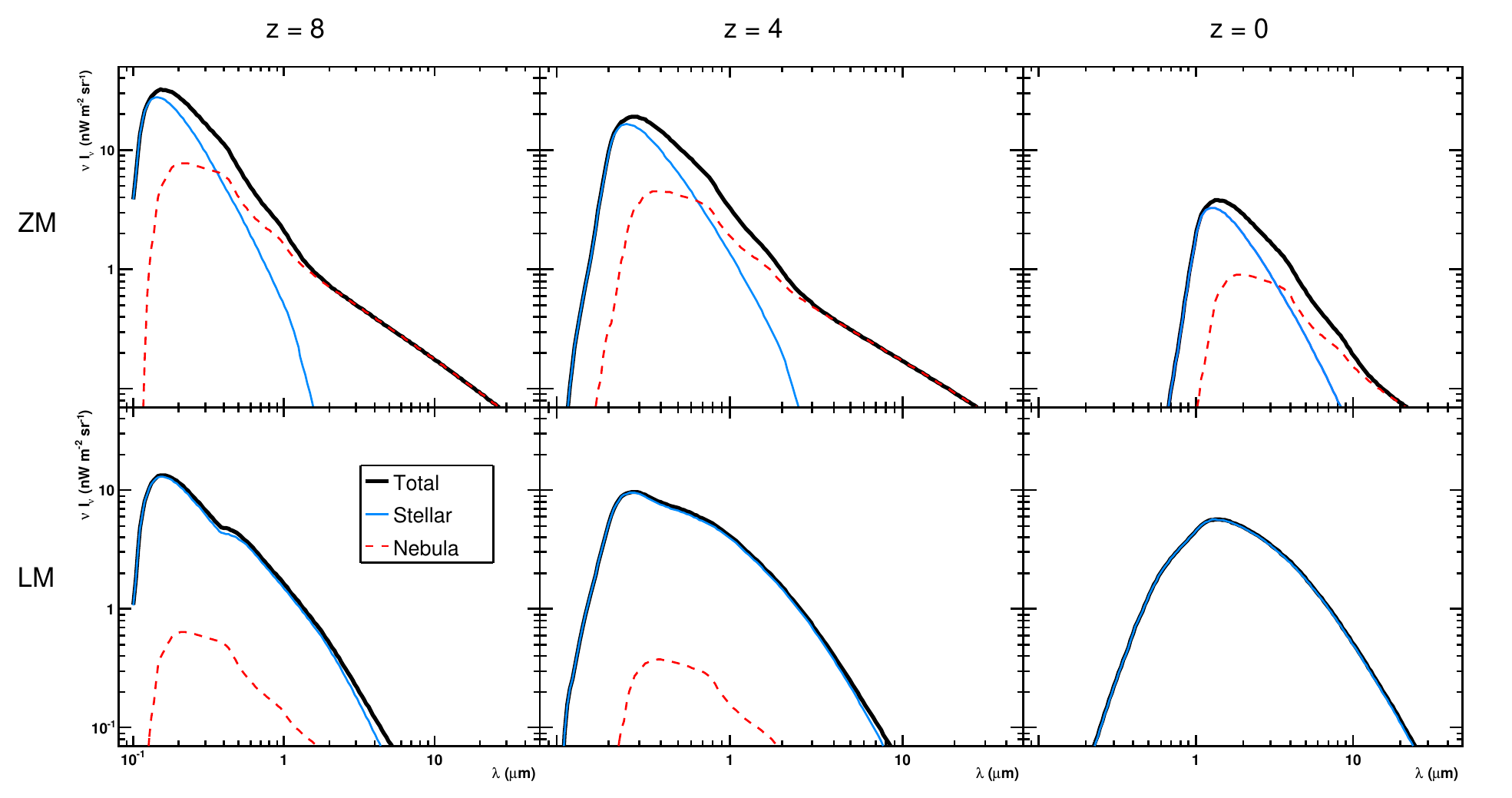}
\caption{Co-moving EBL energy density for ZM stars (top) and LM stars (bottom) for redshift $z = 8$ (left), $z = 4$ (middle), and $z = 0$ (right) and SFR(10, 10, -2). The contribution from the nebula (red dashed) and the stellar (blue) components are shown.}
\label{Fig:EBL_components}
\end{figure*}

The EBL density for the ZM and the LM model for redshifts  $z = 8, 4, 0$ are shown in Fig.~\ref{Fig:EBL_components}  (SFR(10, 10, -2)). For high redshifts, when stars are still being actively formed, the EBL density from ZM stars is factor 2-3 higher than in the LM case. This inverts for low redshifts:  at $z = 0$ the EBL density from LM stars is factor $\sim$2 higher than in the case of ZM stars. The high mass ZM stars have all ended their lifetime and no new photons are produced, while the low mass LM stars are still burning. The contribution from low mass stars also blurs the sharp EBL density drop-off at the ionization edge in the UV, resulting in an overall broader shape and more low wavelength photons than for ZM stars. Consequently, no sharp drop in the EBL at $< 1\,\mu$m is expected for the case of LM stars. For the hot and massive ZM stars the nebula emission makes the dominant contribution to the EBL at the longer wavelengths. For the LM model the contribution from the nebula is negligible. 

These results can be compared with the results from FK06, which only considered stars with mass $> 5$\,M$_\odot$ \footnote{More precisely, stars with lifetimes short enough so that the approximation used by FK06 (no time evolution of emissivity, only averaged quantities) is correct.}: they find that the nebula and line emission are always the dominant contributors to the EBL density, even in the case of low metallicity stars with Salpeter IMF (FK06, Tab.~1). When accounting for the low mass range of the IMF ($< 5$\,M$_\odot$), the emission from old stars dominate the EBL and the nebula contribution is negligible.

\section{Constraining the early star formation}\label{Sec:Constrains}

As discussed in Sec.~\ref{Sec:Introduction}, the EBL can be used as probe for the properties of the first stars. While several authors discussed a possible signature from the PopIII stars in the NIRBE \citep{santos:2002a,salvaterra:2003a,dwek:2005c,madau:2005a,salvaterra:2006a,fernandez:2006a}, we want to focus on the recent limit on the EBL density \citep{aharonian:2006:hess:ebl:nature,mazin:2007a} and what constraints on the properties of the PopIII/LM~PopII stars can be derived from these limits. Two properties will be mainly investigated: (1) the SFR and (2) the influence of the metallicity.

The normalization of the SFR ${\rho}_{0}$ (at  $z = z_{\mathrm{peak}}$) directly translates into the overall normalization of the EBL (see Eq.~\ref{eq:emislambda} and \ref{eq:hinter}). We will set ${\rho}_{0} = 1$ throughout our calculations and later use the scaling relation to derive constraints on ${\rho}_{0}$. The SFR parameters tested are given in Tab.~\ref{Tab:SFR_BPL_Param}. The choice of  $z_{\mathrm{peak}}$ is motivated by the limits on the redshift of reionization derived from WMAP data, which gives a 3\,$\sigma$ lower limit for a sudden reionization of $z_{\mathrm{reion}} > 6.7$ with a best fit of  $z_{\mathrm{reion}} \sim 11 \pm 1.4$ and evidence for an extended process \citep{dunkley:2008a}. A steep slope $\alpha = 10$ corresponds to a sudden stop of the PopIII star formation and negligible contributions at low $z$ while flatter slopes ($\alpha = 4$) give some contribution of the PopIII SFR even at low $z$ as, e.g., predicted by \citet{tornatore:2007a} from numerical simulations. $\beta = 0$ corresponds to an extreme case of a constant SFR (up to $z_{\mathrm{peak}}$) (FK06), while $\beta = -2$ corresponds to a more standard case (e.g. \citealt{dwek:2005c}).

\begin{figure*}[tbp]
\centering
\includegraphics[width=\textwidth]{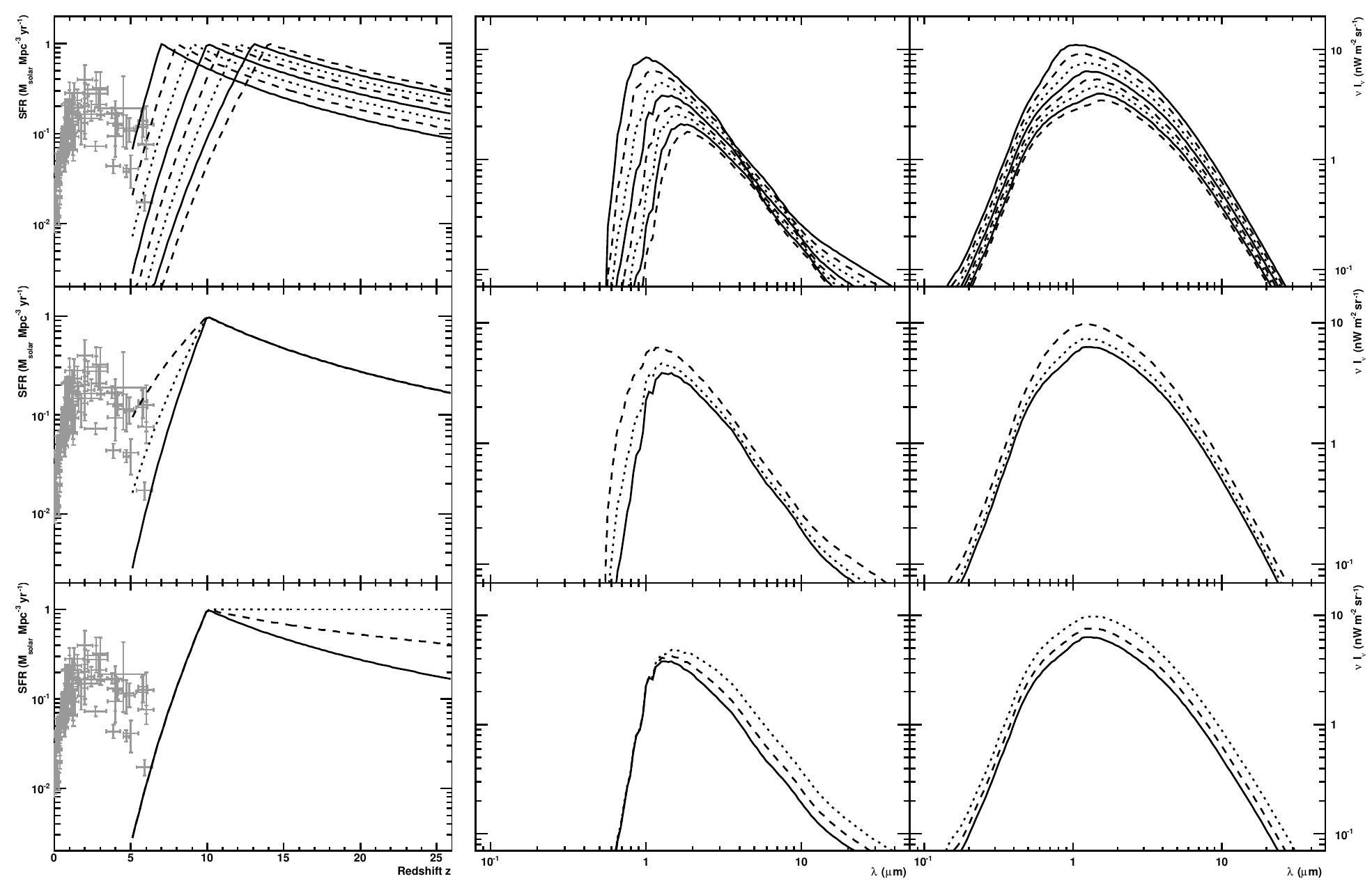}
\caption{Co-moving PopIII SFR (left) and resulting EBL energy density (ZM: middle; LM: right). In each row one parameter of the SFR parameterization is varied. As reference for the low redshift SFR the data collection from \citet{hopkins:2006a} is given (grey markers). \textit{Top:} $z_{\mathrm{peak}}$ varies -- SFR(x, 10, -2). \textit{Middle:} $\alpha$ varies -- SFR(10, x, -2). \textit{Bottom:} $\beta$ varies -- SFR(10, 10, x).}
\label{Fig:EBL_SFR_TumA_BC}
\end{figure*}
 
\begin{table}[tb]
\centering
\begin{tabular}{cc} \hline \hline
Parameter & Values \\ \hline
$z_{\mathrm{peak}}$ & 7, 8, 9, 10, 11, 12, 13, 14\\ 
$\alpha$ &  4, 7, 10 \\
$\beta$ & -2, -1, 0 \\ \hline
\end{tabular}
\caption{Parameter-values of the broken-power-law SFR (Eq.~\ref{Eq:SFR_BPL}). \label{Tab:SFR_BPL_Param}}
\end{table}
 
Figure~\ref{Fig:EBL_SFR_TumA_BC} shows the impact of the different SFR parameters on the resulting EBL energy density. Shifting the peak position $z_{\mathrm{peak}}$ results in a shift in the wavelength of the maximum of the EBL density, and a lower EBL density for higher $z$ (first row). Changing the slope $\alpha$ has a similar effect as changing the peak position, i.e. shifting more of the star formation to smaller/higher $z$ values (second row). The slope $\beta$ changes the EBL density at wavelength greater than the peak in the EBL density (third row). 

The EBL density from LM stars (3rd row) has a broader peak extending to lower wavelengths and a slightly higher overall EBL density than the EBL density from ZM stars (2nd row), mainly due to the longer burning times of the low mass stars. If the first stars have a significant contribution to the overall EBL in excess of the contribution from PopII stars, the detection of a break in the EBL from UV to NIR could be interpreted as an indication for massive ZM stars with short lifetimes (e.g. \citealt{santos:2002a}). Furthermore, the EBL from LM stars shows a different redshift dependency (Fig.~\ref{Fig:EBL_components}), which in principle will result in a different absorption signature for high redshift VHE sources. Detecting such a signature will be a difficult task: high precision measurements of VHE spectra for high redshift are needed and more important, the PopIII contribution to the EBL density has to be in excess of the contribution from second generation of stars (see Sec.~\ref{SubSec:CutOffInHighZ} for a detailed discussion).

\subsection{Initial mass function}\label{SubSec:ConstraintsIMF}

\begin{figure}[tbp]
\centering
\includegraphics[width=0.5\textwidth]{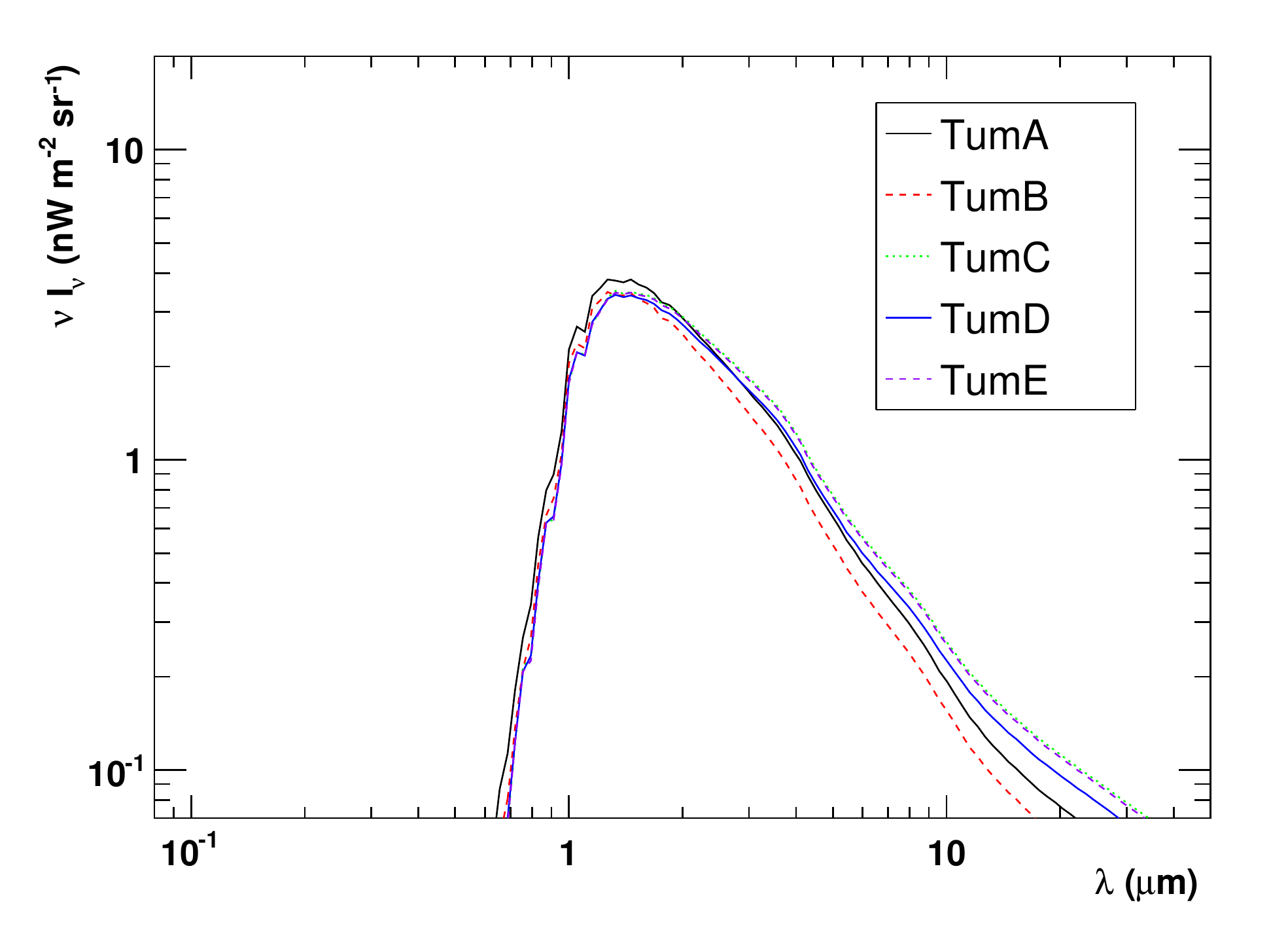}
\caption{Resulting EBL energy density for different IMFs (Tumlinson A, B, C, D, and E) for SFR(10, 10, -2). Given the similarities between the EBL densities, with only small differences at longer wavelengths, we will use the TumA case for further calculations.}
\label{Fig:EBL_TumX}
\end{figure}
 
The zero metallicity SPS from \citet{tumlinson:2006a} are calculated for several different IMF cases shown in Fig.~\ref{Fig:TumIMFs}. Figure~\ref{Fig:EBL_TumX} shows the EBL density resulting for the different IMFs TumA-E using the same SFR(10, 10, -2). For the wavelength range of interest (UV to NIR; $\leqslant 4\,\mu$m), the differences are very small (order $\leqslant 10$\%). To differentiate between such subtle differences the EBL density has to be resolved on the same level. As discussed below (Sec.~\ref{SubSec:VHE_Spectra}), current EBL limits constrain the EBL density in this wavelength range in the order of  $\leqslant 50$\% at best, so for further calculation we will only consider the TumA IMF case (somewhat average EBL) for the ZM stars.

\subsection{Constraints from HE/VHE observations}

\subsubsection*{VHE spectra}\label{SubSec:VHE_Spectra}

\begin{figure}[tbp]
\centering
\includegraphics[width=0.5\textwidth]{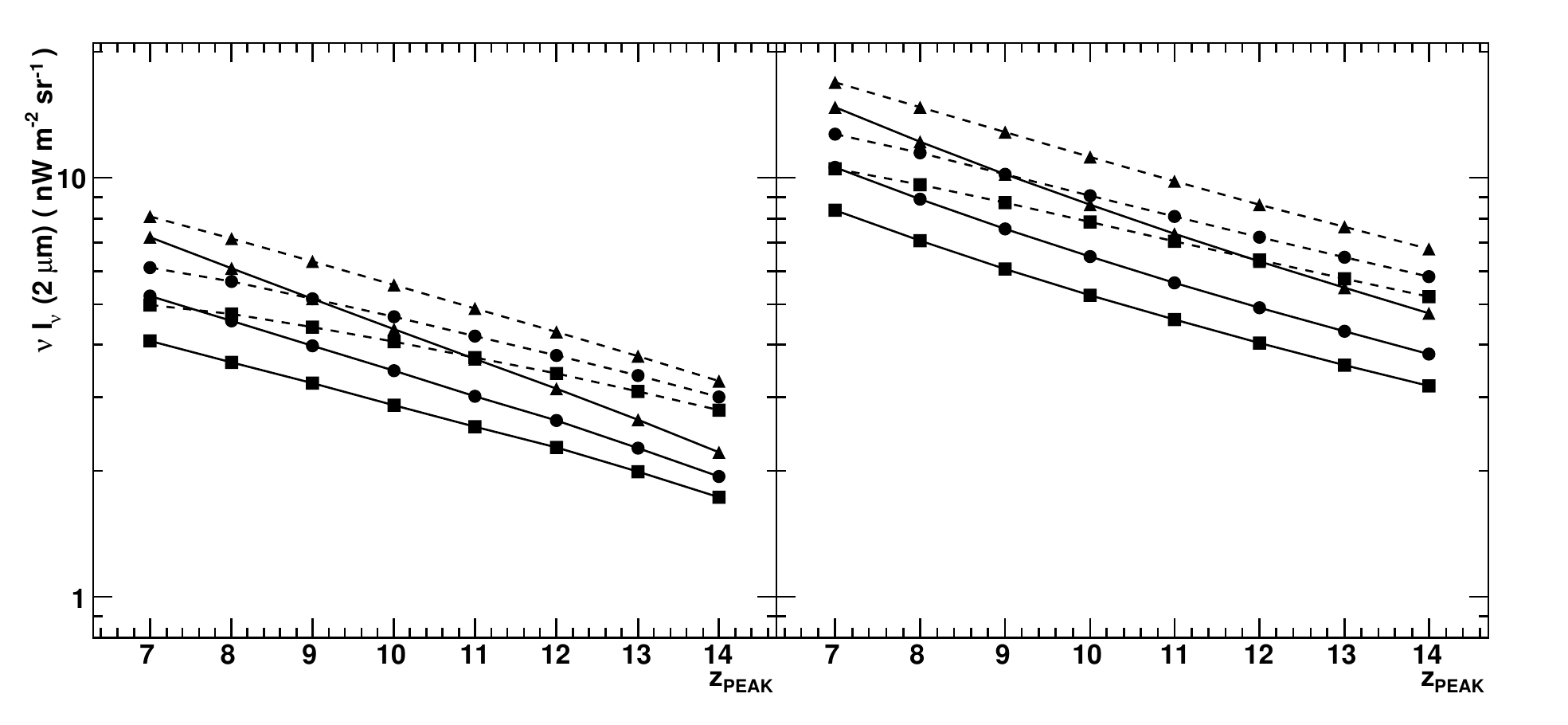}
\caption{EBL density at 2\,$\mu$m ($z = 0$) versus the peak of the SFR $z_{\mathrm{peak}}$ for different stellar models and SFR parameters. \textit{Left:} ZM. \textit{Right:} LM. Other SFR parameters: $\alpha = 10$ (solid lines),  $\alpha = 4$ (dashed lines); $\beta = 0$ (triangles), $\beta = -1$ (circles), $\beta = -2$ (squares). Results for $\alpha = 7$ lie between $\alpha = 10$ and $\alpha = 4$ and are omitted to improve the readability of the plot.}
\label{Fig:EBL2mu_vs_zpeak}
\end{figure}

\begin{figure}[tbp]
\centering
\includegraphics[width=0.5\textwidth]{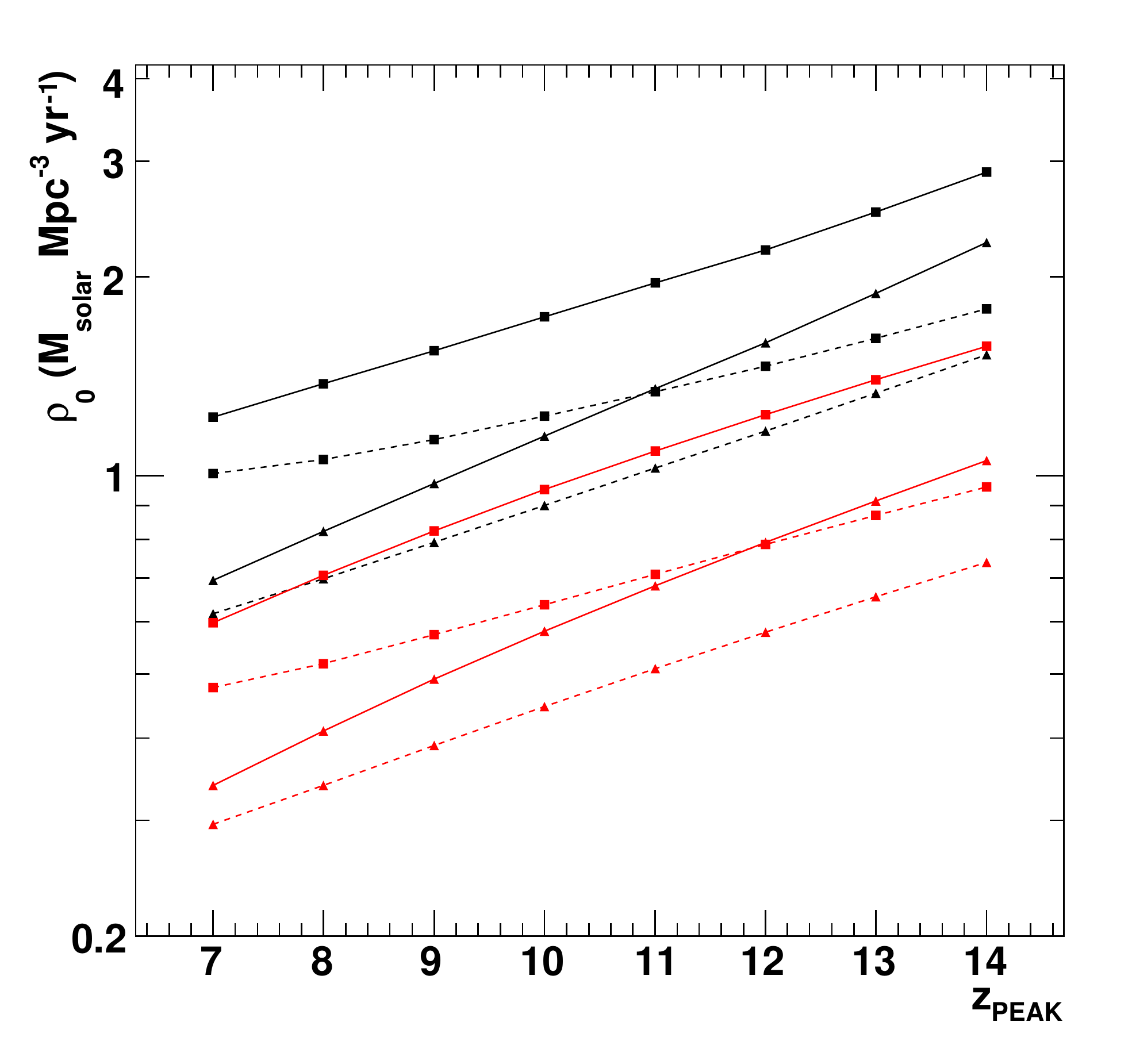}
\caption{Upper limits on the SFR ($\rho_{0}$) derived from the EBL contribution at 2\,$\mu$m assuming a maximum EBL contribution of 5\,nW\,m$^{-2}$\,sr$^{-1}$. Models: ZM (black lines and markers), LM (red/grey lines and markers). SFR parameters: $\alpha = 10$ (solid lines),  $\alpha = 4$ (dashed lines);  $\beta = 0$ (triangles), $\beta = -2$ (squares).}
\label{Fig:SFRmaxnorm_vs_zpeak}
\end{figure}

The contribution from a redshifted early stellar population to the EBL density is expected to peak in the wavelength range around $1 - 2 \, \mu$m. Recently, \citet{aharonian:2006:hess:ebl:nature} derived upper limits of  $\nu I_\nu \lesssim (14 \pm 4)$\,nW\,m$^{-2}$\,sr$^{-1}$ on the EBL density in this wavelength range, utilizing the hard VHE spectrum of the blazar 1ES~1101-232 (located at a redshift $z = 0.186$) and certain assumptions about the source physics (see also \citealt{katarzynski:2006a,stecker:2007a,aharonian:2008a} for some caveats). This limit has been confirmed by the detection of a second source of VHE $\gamma$-rays with similar properties 1ES~0347-121 ($z =0.188$) \citep{aharonian:2007:hess:1es0347} and was shown to be largely independent of the exact shape of the EBL density \citep{mazin:2007a}. \citet{madau:2000a} derive a strict lower limit on the EBL density of $\gtrsim (9.0 - 9.7^{+3.0}_{-1.9})$\,nW\,m$^{-2}$\,sr$^{-1}$ from deep source counts (not corrected for completeness) in the wavelength range $1 - 2\,\mu$m. \citet{totani:2001a} estimated the EBL density contribution from resolved galaxies to be $(10.1 - 12.8)$\,nW\,m$^{-2}$\,sr$^{-1}$ at 1.25\,$\mu$m and  $(7.8 - 10.2)$\,nW\,m$^{-2}$\,sr$^{-1}$ at 2.2\,$\mu$m, accounting for missed galaxies due to selection effect.

In this paper we will adopt a limit on the EBL contribution from the PopIII/LM PopII stars of $\sim 5 $\,nW\,m$^{-2}$\,sr$^{-1}$ at $1 - 2 \, \mu$m.\footnote{This corresponds to the \textit{best guess}, not the extreme limit, i.e. using the outer error range, which is of order $\sim 9 $\,nW\,m$^{-2}$\,sr$^{-1}$.} Comparing this maximum contribution with the EBL density value at 2\,$\mu$m $F_{\mathrm{EBL}}(2\,{\mu}\mathrm{m})$, as calculated from our model for a specific set of parameters, an upper limit on the normalization of the SFR  ${\rho}_{0}$ can be derived:

\begin{equation}
{\rho}_{0} <  5 \,\mathrm{nW}\,\mathrm{m}^{-2}\,\mathrm{sr}^{-1}/F_{\mathrm{EBL}}(2\,{\mu}\mathrm{m})
\end{equation}
whereas $F_{\mathrm{EBL}}(2\,{\mu}\mathrm{m})$ is calculated using ${\rho}_{0} = 1$\,M$_\odot$\,Mpc$^{-3}$\,yr$^{-1}$.

The resulting EBL densities at 2\,$\mu$m for different SFR parameter-sets and metallicities are shown in Fig.~\ref{Fig:EBL2mu_vs_zpeak}. The EBL contribution for the ZM model range from 1.5 to 8.5\,nW\,m$^{-2}$\,sr$^{-1}$, for the LM model from 3 to 15\,nW\,m$^{-2}$\,sr$^{-1}$. Converting these EBL contributions to limits on the normalization of the SFR, limits from 0.3 to 3\,M$_\odot$\,Mpc$^{-3}$\,yr$^{-1}$ are derived (Fig.~\ref{Fig:SFRmaxnorm_vs_zpeak}). LM models result in an overall factor $\sim$2 stronger constraints due to the higher EBL contribution. The different SFR parameters result in a similar spread in the limit of factor 2-3. The limits on the SFR directly scale with the limit on the EBL density contribution at 2\,$\mu$m: if e.g. the EBL limit is lowered from 5 to 1\,nW\,m$^{-2}$\,s$^{-1}$ the corresponding SFR limit is also lowered by a factor of 5.

\subsubsection*{Cut-off in high redshift sources}\label{SubSec:CutOffInHighZ}

\begin{figure*}[tbp]
\centering
\includegraphics[width=0.9\textwidth]{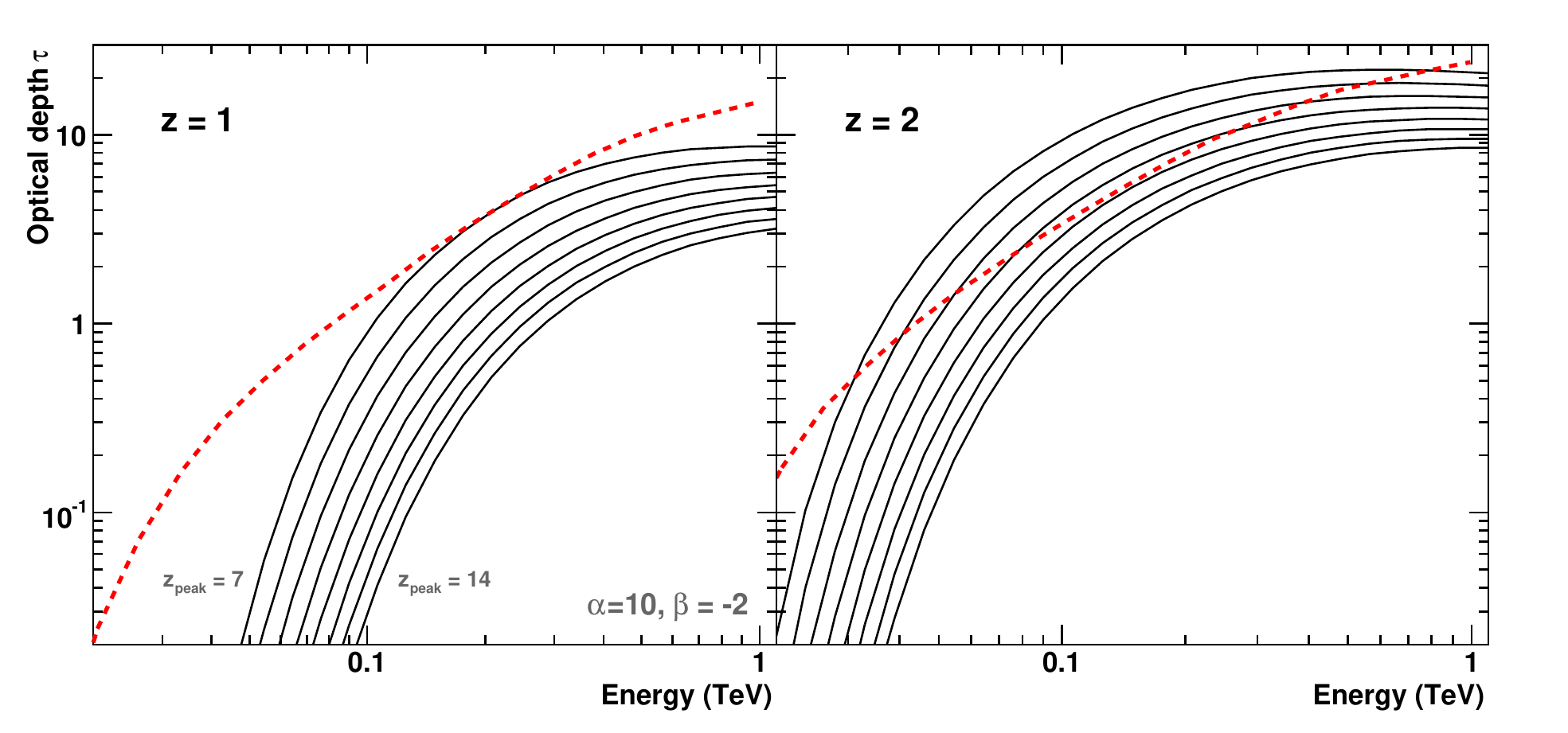}
\caption{Optical depth for VHE $\gamma$-rays derived for the ZM model with SFR(7 - 14, 10, -2) (black solid lines) versus the optical depth derived for a PopII EBL model (low model from \citealt{kneiske:2002a,kneiske:2004a}; red dashed line) for a source at redshift $z = 1$ (left) and $z =2$ (right).}
\label{Fig:tau_pop3_vs_tau_pop2}
\end{figure*}
 
For distant source ($z \gtrsim 1$) of HE/VHE $\gamma$-rays, the attenuation from the EBL results in a strong suppression (cut-off) at energies $> 10 - 100$\,GeV. The cut-off energy shifts with redshift, which in principle could provide a measurement of the EBL density \citep{fazio:1970a}. The cut-off is expected at energies $<$100\,GeV, which will be probed by the Large Area Telescope (LAT) instrument on the recently launched FERMI satellite (energy range: $\sim$20\,MeV -- 300\,GeV).

The main challenge in determining such a cut-off is to distinguish it from source intrinsic effects like e.g. insufficient acceleration of particles to high energies or internal absorption, which could also lead to a cut-off at high energies. Source intrinsic absorption can even lead to a redshift dependent absorption signature, mimicking the z dependent EBL attenuation \citep{reimer:2007a}. Furthermore, energies $\sim$100\,GeV are at the outer edge of the LAT sensitivity band, with the sensitivity decrease by more than one order of magnitude in comparison with the best sensitivity around 1\,GeV\footnote{http://www-glast.slac.stanford.edu/software/IS/\\glast\_lat\_performance.htm}.

Detecting an absorption signature from the PopIII/LM PopII stars is even more challenging: in addition to source intrinsic effects, the absorption from the PopII EBL acts as foreground, from which the absorption from the PopIII EBL has to be differentiated. Figure~\ref{Fig:tau_pop3_vs_tau_pop2} shows the optical depth derived for the ZM PopIII model in comparison to the optical depth from the low PopII model from \citet{kneiske:2002a,kneiske:2004a} for a source at redshift $z =1$ and $z =2 $.\footnote{For details on the calculation of the optical depth see e.g. \citealt{mazin:2007a}.} For redshift $z = 1$, the optical depth resulting from the PopII EBL dominates over the PopIII contribution (with $\rho_0 = 1$, which is of order of the limit derived in the previous section). For higher redshifts the situation is different: the PopII SFR rises steeply with redshift with a plateau or peak expected in the redshift range $z = 1 - 2$, so the main PopII EBL contribution is building up over the redshift range $z = 0 - 2$. The PopIII will not add new photons to the EBL in this redshift range, so the co-moving EBL density contribution from PopIII stars is constant and just scales with the cosmological expansion/contraction. Consequently, at redshift $z \gtrsim 2$ the optical depth resulting from PopIII EBL can dominate the total attenuation at energies $\geqslant$30\,GeV (Fig.~\ref{Fig:tau_pop3_vs_tau_pop2} right). For a source at redshift $z \sim 2$ the optical depth resulting from the PopII EBL is $\geqslant 1$, so to distinguish between PopIII and PopII contribution, one would have to differentiate between optical depths of e.g. 1 and 10. Note that the PopII EBL model could underestimate the EBL density in the UV-O range, since the contribution from AGN is not included in the calculation. Thus the background resulting from other contributors to the EBL could be even higher.

In case of the LM scenarios the resulting optical depth is higher than for the ZM models, due to the higher EBL density at $\lambda <  0.5 \mu$m (Fig.~\ref{Fig:EBL_SFR_TumA_BC}), but this will not change the fundamental challenge of determining the exact shape of a steep cut-off from a low statistic measurement.

\section{Discussion}\label{Sec:Discussion}

\begin{figure}[t,b]
\centering
\includegraphics[width=0.5\textwidth]{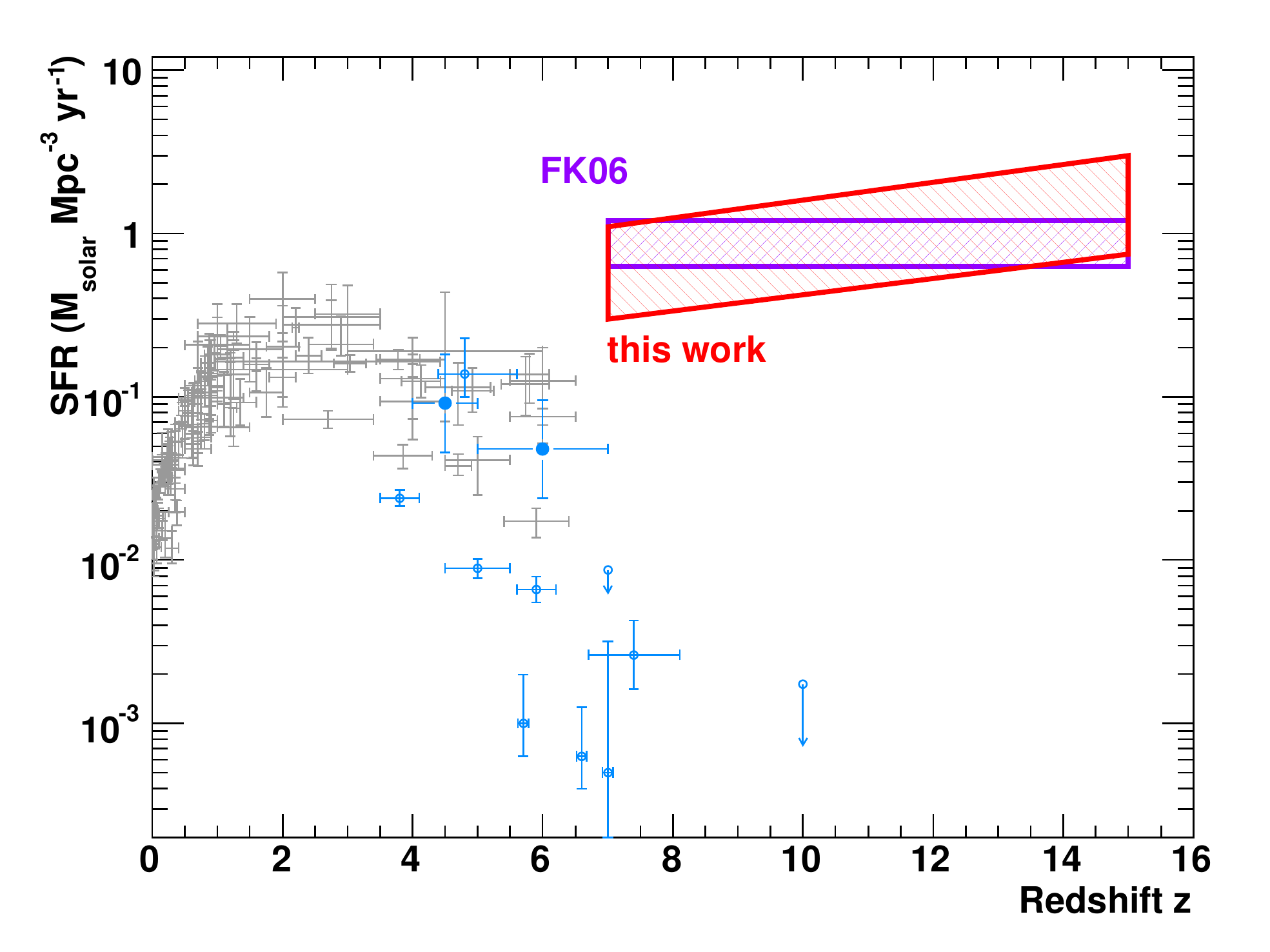}
\caption{Limits on the co-moving SFR of PopIII stars derived in this paper in comparison to other limits and measurements of the star formation rate. The purple striped region indicates the range given in FK06 (see text for details). Blue open markers are from the data collection from \citet{yuksel:2008a}, while blue filled markers at $z = 4.5$ and $z = 6$ are the data points derived in the same paper from GRB measurements. As reference for the low redshift SFR the data collection from \citet{hopkins:2006a} is given (grey markers).}
\label{Fig:SFR_results}
\end{figure}
 
Figure~\ref{Fig:SFR_results} shows the limit on the PopIII SFR derived in this paper in comparison with other measurements and limits on the SFR. As argued in Sec.~\ref{SubSec:SFR}, the direct measurements (from e.g. source counts) are not directly comparable with the limits derived from an integrated property like the EBL, since they most likely sample different contribution to the total SFR. In this respect, they have to be considered lower limits on the total SFR. The direct measurements at $z \sim 7$ lie two to three orders of magnitude below our limits, while recent determination of the SFR at $z = 3 - 6$ give $~0.1$\,M$_\odot$\,Mpc$^{-3}$\,yr$^{-1}$ \citep{yuksel:2008a}. \citet{henry:2008a} reported a candidate for a Lyman break galaxy at $z = 9$. If confirmed, this could imply that the SFR at $z \sim 9$ does not evolve strongly and is roughly on the same level as at $z = 3$ \citep{henry:2008a}, which would bring our limits in range (factor 5-10) of the direct measurements.

Other results on the SFR of PopIII stars come from numerical simulations: \citet{bromm:2002b} derive a peak SFR of 0.3-0.6\,M$_\odot$\,Mpc$^{-3}$\,yr$^{-1}$ depending on the dominant cooling mechanism. \citet{tornatore:2007a} find in their simulation low SFRs with peaks of 10$^{-5}$ and 10$^{-4}$\,M$_\odot$\,Mpc$^{-3}$\,yr$^{-1}$ at $z \approx 6$ for PopIII and PopII stars, respectively.

Valuable constraints on the properties of PopIII star have also been derived from reionization studies. \citet{choudhury:2006a} used a self-consistent reionization model together with experimental results to derive a best fit model of the reionization history.
Their best fit model, which predicts negligible source counts at $z > 10$, has a peak SFR for the PopIII stars of order $10^{-3}$\,M$_\odot$\,Mpc$^{-3}$\,yr$^{-1}$ ($z \sim 12$) and is always dominated by LM PopII stars (PopII SFR rising to $0.1$\,M$_\odot$\,Mpc$^{-3}$\,yr$^{-1}$ at $z = 7$, smoothly connecting with the low redshift PopII SFR). However, if some of the high redshift source candidates from deep surveys (see Sec.~\ref{SubSec:SFR}) would turn out to be valid $z \approx 10$ sources, requiring a higher SFR at $z = 10$, the model would be strongly constrained by the limits on the Ly $\alpha$ optical depth, possibly allowing to pin down how reionization occurred.
\citet{greif:2006a} investigated the reionization history with a semi-analytic model utilizing two distinct ZM PopIII star populations\footnote{Classical high mass $\gtrsim 100$\,M$_\odot$ PopIII stars from clouds with molecular hydrogen and lower mass  $\gtrsim 10$\,M$_\odot $PopII.5 stars, formed in clouds with HD cooling.}. They investigated the effect of several feedback mechanisms (Lymann-Werner, photoheating, chemical feedback) on the SFR of PopIII stars and found the peak SFR of PopIII in the order of  $10^{-4} - 10^{-3}$\,M$_\odot$\,Mpc$^{-3}$\,yr$^{-1}$. Again, the overall SFR in the redshift range $z = 10 - 15$ is dominated by (LM) PopII stars ($\sim 10^{-2} - 10^{-1}$\,M$_\odot$\,Mpc$^{-3}$\,yr$^{-1}$ at $z = 5 - 15$). \citet{haiman:2006a} found evidence for a suppression of the SFR in high-redshift minihalos, utilizing the WMAP constraints on the optical depth to electron scattering.
For zero metallicity PopIII stars the limits on the SFR derived in this paper are 1 to 5 orders of magnitude above the predictions from reionization models. For LM PopII the limits are still above the predictions, but only a factor 3 to 10.

FK06 derived limits on the PopIII star formation rate using a simplified EBL model, assuming a PopIII EBL contribution of 2 - 50\,nW\,m$^{-2}$\,sr$^{-1}$. We convert their result (FK06, Tab.~1) using a maximum PopIII EBL of  5\,nW\,m$^{-2}$\,sr$^{-1}$. The resulting limit on the PopIII SFR is 0.63 - 1.2\,M$_\odot$\,Mpc$^{-3}$\,yr$^{-1}$ (Fig.~\ref{Fig:SFR_results}, purple striped box), which is in the range of the limit derived in this paper. Note that these results are not directly comparable with our results since (1) FK06 used a simplified model for the stellar emission, which does not account for the temporal evolution of the emission. While for ZM stars with top-heavy IMFs this is a valid approximation (high mass corresponds to short lifetime), for LM stars with Salpeter IMF the EBL contribution from low-mass and therefore long-lived stars changes the shape of the EBL and increase the overall EBL density  (see Sec.~\ref{SubSec:ResultingEBL}). (2) FK06 only consider a constant SFR similar to our case SFR(x, 10 -2). Considering different SFRs introduces a redshift dependency of the limit and a spread of factor $\sim 5$.

\begin{table}[tb]
\centering
\begin{tabular}{ccccccc}\hline \hline
 & TumA &TumB & TumC & TumD & TumE & Salpeter \\ \hline
$f_{\mathrm{rem}}$ & 0.27 & 0.12 & 0.56 & 0.37 & 0.43 & 0.6\\
$f_{\mathrm{metal}}$ & 0.12 & 0.15 & 0.07 & 0.11 & 0.27 & $< 0.047$ \\ \hline 
\end{tabular}
\caption{Remnant mass fraction $f_{\mathrm{rem}}$ (Eq.~\ref{eq:f_rem}) and metal mass fraction  $f_{\mathrm{metal}}$ for different IMFs. For $f_{\mathrm{rem}}$ for the Salpeter IMF results from  \citet{nagamine:2006a} have been used. \label{Tab:BaryonsMetalsInStars}}
\end{table}
 
An upper limit on the baryonic mass density bound in PopIII stars $ M_{\ast}$ can be derived by integrating the SFR over time
\begin{equation}
 M_{\ast}
  <  M_{\ast}^{\mathrm{lim}} = \int_{0}^{t_{\mathrm{start}}} {\rho}_{\ast}(t) \, ,
\label{eq:IntSFR}
\end{equation}
which can be compared with estimates for the total density of baryons locked up in stars of 6\%$\pm$2\% $\Omega_{\mathrm{b}}$ \citep{fukugita:2004a}, with $\Omega_{\mathrm{b}}$ being the baryonic density of the universe in units of the critical density. For our limiting SFRs we derive $M_{\ast}^{\mathrm{lim}} \approx$5\%\,$\Omega_{\mathrm{b}}$ for the LM and $M_{\ast}^{\mathrm{lim}} \approx$10\%\,$\Omega_{\mathrm{b}}$ for the ZM stars. Since not all baryons remain locked up in stars (they get ejected by supernova explosions or stellar winds) this number has to be corrected for baryons leaving the stars. \citet{nagamine:2006a} calculated the fraction of amount of recycled gas to the total amount of gas initially converted to stars to be $f_{\mathrm{rec}}= 0.32$ for Salpeter IMF and solar metallicity. Adopting a remnant mass fraction of 0.6 of the initial mass the value for LM stars is lowered to $M_{\ast} \sim $3\%\,$\Omega_{\mathrm{b}}$. For the ZM stars with lifetimes $\sim 10^6$\,yr it is reasonable to assume that all ZM stars created at redshifts $z  > 5$ have reached their final state. We therefore estimate the remnant mass fraction $f_{\mathrm{rem}}$ using the final state masses for ZM stars presented in  Fig.~2 of \citet{heger:2002a}:
\begin{equation}
f_{\mathrm{rem}} = \frac{\int_{m_{\mathrm{min}}}^{m_{\mathrm{max}}} N(m)\,M_{\mathrm{rem}}(m)\,dm}{\int_{m_{\mathrm{min}}}^{m_{\mathrm{max}}} N(m)\,m\,dm}
\label{eq:f_rem}
\end{equation}
with $N(m)$ being the IMF and $M_{\mathrm{rem}}(m)$ the remnant masses. The derived fractions range from 0.12 to 0.56 depending on the IMF (Tab.~\ref{Tab:BaryonsMetalsInStars}), which results in a baryon density locked up in ZM stars of $\sim1.2-5.6\%\,\Omega_{\mathrm{b}}$. The final states of ZM stars are not fully understood, so these numbers should be considered estimates not precise calculations. However, these estimates demonstrate that our SFR limits are below the range what could be excluded from the limits on the density of baryons in stars.

Similar calculations can be performed for the metal enrichment from the PopIII/LM PopII stars. The metal mass fraction $f_{\mathrm{metal}}$ can be calculated from Eq.~\ref{eq:f_rem} by replacing $M_{\mathrm{rem}}(m)$ with $M_{\mathrm{metal}}(m)$, the metal yield produced by a star of mass m. Here it is again implicitly assumed that all star have reached their final stage. For ZM stars this is a valid assumption, but for the case of LM stars this calculation results in an upper limit on the metal density produced by the stars. We adopt metal yields from \citet{heger:2002a} ($Z = 0$; $M > 120 M_\odot$), \citet{portinari:1998a} ($Z = 0.0004$; $5\,M_\odot < M < 120\,M_\odot$) and \cite{marigo:2001a} ( $Z = 0.004$; $0.8\,M_\odot < M < 5\,M_\odot$). The results are summarized in Tab.~\ref{Tab:BaryonsMetalsInStars}: LM stars produce metal densities of $< 0.2\%\,\Omega_{\mathrm{b}}$, which is well below todays metal density of $\sim 2\%\,\Omega_{\mathrm{b}}$. ZM stars produce metal densities of $\sim 0.7-2.7\%\,\Omega_{\mathrm{b}}$ depending on the IMF. Again, we stress that these are estimates, since there are large uncertainties in the final states of the ZM stars, especially in the high mass range. Still, ZM stars with very heavy IMFs (e.g. the Tum-E case) seem to overproduce the metal density, if a SFR on the level of our limits is assumed.

\section{Summary \& Conclusions}\label{Sec:Summary}

We investigate how limits on an integrated present-day observable, the EBL, can be used to constrain the parameters of the early stars. A detailed model for the PopIII/LM PopII star emission from a large range of different scenarios is used to calculate the evolving EBL from these stars, taking into account the time evolution of the emissivity and the emission from reprocessed ionizing photons (nebula). Recent limits on the EBL density derived by \citet{aharonian:2006:hess:ebl:nature} from the detection of hard VHE $\gamma$-ray spectra from distant sources together with lower limits from source counts \citep{madau:2000a,totani:2001a} suggest a maximum PopIII EBL contribution of $\sim 5 $\,nW\,m$^{-2}$\,sr$^{-1}$ at $1-2\,\mu$m. Comparing this contribution with our model calculations, a limit on the co-moving SFR of PopIII stars of 0.3 to 3\,M$_\odot$\,Mpc$^{-3}$\,yr$^{-1}$  is derived for the redshift range $7 - 14$. This limit depends on the redshift, on the exact shape of the SFR and on the assumed scenario for the early star formation: if the early star formation is dominated by second generation stars with low metallicity, the limit is factor two lower than in the case of zero metallicity stars. The SFR limit directly scales with the assumed PopIII EBL contribution, e.g., if the EBL limit is lowered by a factor 2 the corresponding SFR limit is also lowered by the same factor.

The SFR at redshift $> 5$ is difficult to access via direct observations. A few measurements and limits exist, generally favoring a lower SFR in the range of 10$^{-3}$ to 10$^{-2}$\,M$_\odot$\,Mpc$^{-3}$\,yr$^{-1}$, but the spread and uncertainties are large (Fig.~\ref{Fig:SFR_results}). Recent measurements of the SFR in the redshift range $z = 3 - 6$ favor a flat (or even increasing) SFR of the order of 0.1\,M$_\odot$\,Mpc$^{-3}$ \citep{yuksel:2008a,faucher-giguere:2008a:astro-ph}, which is in the range of our best limit (0.3\,M$_\odot$\,Mpc$^{-3}$\,yr$^{-1}$ at $z = 7$ for LM, SFR(7, 4, 0)).
Stringent constraints on the properties of the first stars come from reionization studies, which combine complex semi-analytical modeling with limits on the reionization history.
Predictions for the peak SFR for PopIII and (LM) PopII  are in the range of $10^{-4} - 10^{-3}$\,M$_\odot$\,Mpc$^{-3}$\,yr$^{-1}$ and $\sim 10^{-2} - 10^{-1}$\,M$_\odot$\,Mpc$^{-3}$\,yr$^{-1}$, respectively \citep{choudhury:2006a, greif:2006a}.\footnote{A higher SFR is also possible, see Sect.~\ref{Sec:Discussion}.} While for PopIII stars the limits derived in this paper are 1 to 5 orders of magnitude above the SFR expected from the best fit models from these studies, for (LM) PopII stars the limits are close (factor 3 to 10) to these predictions.

Pair-creation of VHE photons from distant sources ($z > 1$) with the low energy photons from the EBL results in a sharp cut-off in energy spectra $\gtrsim 30$\,GeV, which should be detected by the FERMI experiment. To derive constraints on the PopIII/LM PopII stars from the detection of such a cut-off is challenging since (a) the photon statistics will likely be low, (b) attenuation from the PopIII/LM PopII EBL competes with the attenuation due to the EBL from PopII stars, which is likely the dominant contribution to the total attenuation, and (c) the general problem to discriminate between source intrinsic effects and attenuation from the EBL.

Constraints on the EBL can provide additional insides in the star formation processes of the early universe.
Though the limits are not (yet) strongly constraining, they provide an independent probe for the star formation at redshift $z > 5$.
With the current limits on the EBL in the NIR it is not possible to distinguish between different PopIII IMFs or metallicity scenarios. In the future, the Cherenkov Telescope Array (CTA)\footnote{http://www.cta-observatory.org/} will provide sensitive measurements in the $\sim$20\,GeV to 100\,TeV energy range, which will result in strong constraints on the EBL in a wide wavelength range. Together with direct detections and deep source counts from upcoming satellite and ground-based telescopes this will enable to resolve many of the contributors to the EBL and thereby tighten the limits on the PopIII/LM PopII stars properties derived from the EBL.


\begin{acknowledgements}
We would like to thank Jason Tumlinson for kindly providing us with stellar population spectra for his IMF test cases, Andrew M. Hopkins and Hasan Y\"uksel for sharing their results and data collections of SFR measurements and Laura Portinari for providing us with stellar metal yields. The authors thank Dieter Horns for reading of the manuscript and helpful comments. M.~R. gratefully acknowledge the support by an LEA Fellowship. T.~K.'s research was partially funded by the DFG grant KN 765/1-2. The authors acknowledge the use of the Rechenzentrum Garching. This research has made use of NASA's Astrophysics Data System.
\end{acknowledgements}


\bibliographystyle{aa}
\bibliography{raue_kneiske_mazin_2008}

\end{document}